\documentclass[onecolumn]{aastex631}

\usepackage{amsmath,amsthm,amssymb,amsfonts,xspace,graphicx}
\usepackage{fontawesome}
\usepackage{tikz}
\usetikzlibrary{fit, positioning} 
\usetikzlibrary{shapes, arrows.meta, positioning}
\usepackage{booktabs}
\usepackage{soul}

\shorttitle{Maven: A Multimodal Foundation Model for Supernova Science}
\shortauthors{Helfer, Zhang, et al}

\newcommand{\package}[1]{\textsl{#1}\xspace}

\begin{document}\sloppy\sloppypar\raggedbottom\frenchspacing

\title{Maven: A Multimodal Foundation Model for Supernova Science}

\author[0000-0002-8019-8082]{Gemma Zhang}
\altaffiliation{Both authors contributed equally to this work. Authors may be listed in either order.}
\affiliation{The NSF AI Institute for Artificial Intelligence and Fundamental Interactions}
\affiliation{Department of Physics, Harvard University, Cambridge, MA 02138, USA}

\author[0000-0001-6880-1005]{Thomas Helfer}
\altaffiliation{Both authors contributed equally to this work. Authors may be listed in either order.}
\affiliation{Institute for Advanced Computational Science, Stony Brook University, Stony Brook, NY 11794 USA}

\author[0000-0003-4906-8447]{Alexander~T.~Gagliano}
\affiliation{The NSF AI Institute for Artificial Intelligence and Fundamental Interactions}
\affiliation{Department of Physics, Massachusetts Institute of Technology, Cambridge, MA 02139, USA}
\affiliation{Center for Astrophysics | Harvard \& Smithsonian, 60 Garden Street, MS-16, Cambridge, MA 02138, USA}

\author[0000-0001-9088-7845]{Siddharth Mishra-Sharma}
\altaffiliation{Currently at Anthropic; work performed while at MIT/IAIFI.}
\affiliation{The NSF AI Institute for Artificial Intelligence and Fundamental Interactions}
\affiliation{Center for Theoretical Physics, Massachusetts Institute of Technology, Cambridge, MA 02139, USA}
\affiliation{Department of Physics, Harvard University, Cambridge, MA 02138, USA}

\author[0000-0002-5814-4061]{V.~Ashley Villar}
\affiliation{Center for Astrophysics | Harvard \& Smithsonian, 60 Garden Street, MS-16, Cambridge, MA 02138, USA}
\affiliation{The NSF AI Institute for Artificial Intelligence and Fundamental Interactions}

\email{\href{mailto:yzhang7@g.harvard.edu}{yzhang7@g.harvard.edu} and \href{mailto:thomashelfer@live.de}{thomashelfer@live.de}}

\begin{abstract}\noindent
A common setting in astronomy is the availability of a small number of high-quality observations, and larger amounts of either lower-quality observations or synthetic data from simplified models. Time-domain astrophysics is a canonical example of this imbalance, with the number of supernovae observed photometrically outpacing the number observed spectroscopically by multiple orders of magnitude. At the same time, no data-driven models exist to understand these photometric and spectroscopic observables in a common context. Contrastive learning objectives, which have grown in popularity for aligning distinct data modalities in a shared embedding space, provide a potential solution to extract information from these modalities. We present Maven, the first foundation model for supernova science. To construct Maven, we first pre-train our model to align photometry and spectroscopy from 0.5M synthetic supernovae using a constrastive objective. We then fine-tune the model on 4,702 observed supernovae from the Zwicky Transient Facility. Maven reaches state-of-the-art performance on both classification and redshift estimation, despite the embeddings not being explicitly optimized for these tasks. Through ablation studies, we show that pre-training with synthetic data improves overall performance. In the upcoming era of the Vera C. Rubin Observatory, Maven serves as a Rosetta Stone for leveraging large, unlabeled and multimodal time-domain datasets.
\end{abstract}


\section{Introduction} \label{sec:intro}
The discovery rate of supernovae (SNe) has grown exponentially over the past four decades, thanks in large part to wide-field, untargeted optical surveys (e.g.,  All Sky Automated Survey for SuperNovae (ASAS-SN; \citealt{shappee2014all}), ATLAS (\citealt{tonry2018atlas}), the Zwicky Transient Facility (ZTF; \citealt{bellm2018zwicky}) and the Young Supernova Experiment (YSE; \citealt{jones2021young}). Today, well over ten-thousand SNe are discovered annually. The upcoming Legacy Survey of Space and Time (LSST; \citealt{2019Ivezic_VRO}), conducted by the Vera C. Rubin Observatory, is expected to commence in 2025 and will continue for ten years. LSST will enable the photometric discovery of over one million SNe annually, in addition to millions of other non-SN variable phenomena (including flaring stars and active galactic nuclei). We expect a small fraction of LSST SNe--no more than 1\%--to be observed spectroscopically.

\begin{figure}[t]
  \centering \includegraphics[width=1.0\textwidth]{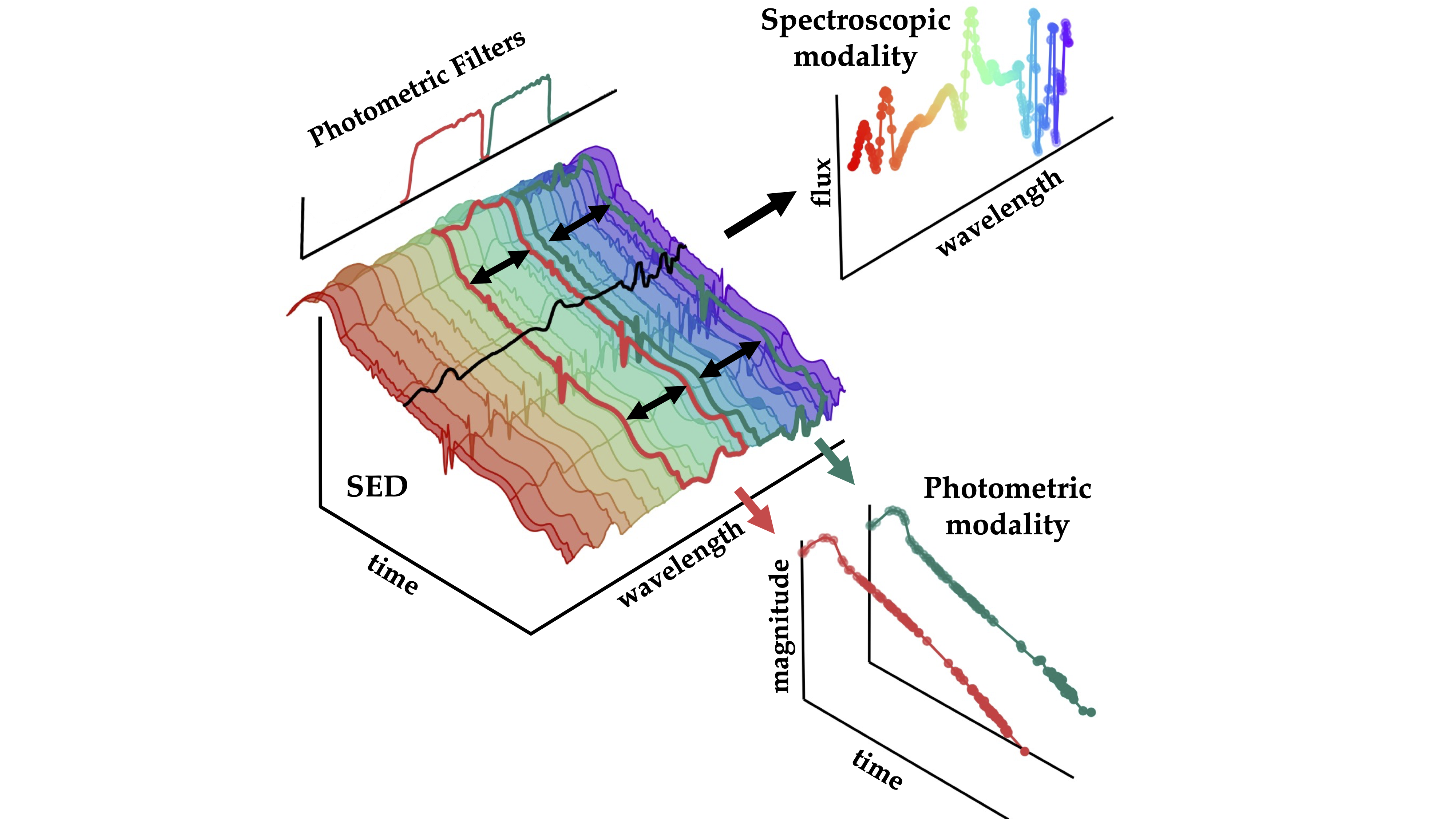} 
  \caption{{\bf Illustration of the mutual information between modalities} considered in this work. Supernovae are characterized by a spectral energy distribution (SED) varying in wavelength and time. Photons are collected through broad-band filters (ZTF-$g$ and ZTF-$r$ with transmission curves shown in the top left) at multiple epochs during explosion to construct multi-band light curves (bottom right).  At a single epoch, supernova light can also dispersed with a spectrograph (SEDM in this work) to obtain a spectrum (top right). Both modalities offer a complementary, but limited, view of an SN's underlying SED.}
  \label{fig:overview_SED}
\end{figure}

SNe can be characterized by a spectral energy distribution (SED), the energy emitted by the event over wavelength and time. Data from this SED at a fixed time is observed as spectroscopy, and for a fixed wavelength range is observed as photometry. This photometry is collected over time to construct a SN's light curve (see Fig.~\ref{fig:overview_SED}). While photometry is easily obtained, spectroscopy is significantly more time-consuming to acquire (long integration times are needed to build up sufficient signal across a spectrograph). This challenge has catalyzed research into techniques to infer the underlying physics of an explosion directly from photometric observations, including the classification of SN types \citep[e.g.,][]{2019Muthukrishna_RAPID,villar2019supernova, 2020Moller_supernnova,2021Boone_parsnip,2023Gagliano_FirstImpressions,2024Rehemtulla_BTSbot} and inference of SN redshifts \citep{2023Mitra_photozs,2023Qu_photozs}. In this context, supervised machine learning has dominated the training of models for the classification of SN types and the estimation of SN redshift. The labels used in the supervised training scenario must be first extracted from spectra, demanding large spectroscopic datasets for sufficient model performance. 
To overcome this issue, researchers have begun to explore self-supervised learning to leverage the structure of unlabeled photometric datasets, by training a feature extraction network and generating a low-dimensional latent space \citep{2012Richards_SelfSupervised,2020Villar_Superraenn}. The learned latent space can then be used to classify events using supervised methods.

Self-supervised representation learning for time-domain astrophysics is appealing for multiple reasons. Pre-trained models have been shown to produce latent data representations that are more robust against distribution shifts than their supervised counterparts \citep{shi2022robust,2022Goyal_VisionModels}. Distribution shift is a common obstacle when applying models trained on bright, spectroscopically-confirmed low-redshift transients to fainter, more distant phenomena that are underrepresented in the training data. Self-supervised learning may also be less sensitive to the class imbalances observed in transient science \citep{2020Yang_ClassImbalance}: labeled SN samples are dominated by type Ia SNe due to their high luminosities relative to other classes. The generalizability of learned representations \citep{2021Kim_SelfReg,2022Ericsson_SSL} also offers the potential for using a pre-trained model for multiple inference tasks and across diverse time-domain surveys, with only minimal fine-tuning.

Contrastive learning has emerged as an effective pre-training objective for combining multiple data modalities. \cite{radford2021learning} present an embedding scheme called Contrastive Language–Image Pre-training (CLIP) for aligning natural language and associated images in a shared latent space. Following this example, domain-specific `foundation models' are beginning to appear in the literature. \cite{parker2024astroclip} recently introduced a cross-modal foundation model using galaxy spectroscopy and images. After independently embedding galaxy images and spectra into low-dimensional latent spaces, they use contrastive training to align the embeddings into a joint latent space. They find that such a model can achieve state-of-the-art performance on the inference of various physical properties (including redshift, mass, and age). Similarly, \cite{2024Slijepcevic_RadioGalaxies} leveraged contrastive learning with instance differentiation, and created a foundational model for radio galaxies by augmenting and aligning unimodal data instances via simple transforms such as rotations. Their resulting model is able to perform accurate morphological classification with fewer labels than supervised methods.

Here, we present Maven, the first multimodal foundation model for SNe. In contrast to previous models for SN classification and redshift inference, our model is constructed using spectroscopic and photometric information simultaneously. Motivated by previous work in synthetic pre-training, we first train Maven by aligning simulated light curve-spectrum pairs via contrastive learning, and fine-tune it on a small sample of observed data using the same approach. Our final model achieves state-of-the-art performance on multiple downstream tasks. We also train a model with only observed data, called Maven-lite, to quantify the impact of synthetic pre-training. Though we limit our analysis to classification and redshift (two popular inference tasks in SN science), the model is a milestone toward general-purpose training for a range of downstream tasks.

Our paper is organized as follows. In Section~\ref{sec:data}, we describe the simulated and observed data used in this work. In Section~\ref{sec:methods}, we describe the architectures of our photometric and spectroscopic encoder models, the contrastive learning objective used to pre-train and fine-tune Maven, and the downstream tasks we use to evaluate Maven's performance. We present our results in Section~\ref{sec:results}, and compare our model to baseline transformer models optimized explicitly for the explored tasks. We further compare our results to existing transformer-based models from the literature. We conclude by discussing the value of contrastive pre-training in astronomy and potential future research directions in Sections~\ref{sec:discussion} and \ref{sec:conclusion}.

\section{Datasets and Simulations}
\label{sec:data}

In this study, we utilize two datasets: a simulated dataset for pre-training and a dataset of observations for subsequent fine-tuning and validation\footnote{All data are available at \url{https://huggingface.co/datasets/thelfer/multimodal_supernovae}}. We describe the details of each below. 

\subsection{Simulating Supernovae with the SNANA Simulation Code}\label{subsec:sims}
We generate synthetic SN samples using the SNANA simulation code \citep{2009Kessler_SNANA}. SNANA mimics the observing process beginning from a rest-frame spectral energy distribution (SED) of an astrophysical transient. A volumetric rate is chosen and the sky is populated at random with transients. A survey strategy, detection efficiency, and the survey’s estimated noise properties (zeropoint and sky noise) are provided to construct synthetic observations.

We simulate observations of the Zwicky Transient Facility \citep{bellm2018zwicky} using the framework described in \citet{2023Aleo_YSEDR1}, which approximately matches the redshift distribution of the SNe in our observed sample (described in the following section \ref{subsec:ZTFBTS}). We simulate 500,000 total events evenly split between five different SN classes, using SED models from the Photometric LSST Astronomical Time-Series Classification Challenge \citep{2019Kessler_PLAsTiCC}: SNe~Ia (using the \verb|SALT2| model; \citealt{2007Guy_SALT2}); SNe~Ib/c (\verb|SNIbc-Templates|; \citealt{kessler2010results}); SLSNe-I (using the model \verb|SLSNI-MOSFIT|; \citealt{villar2017theoretical}); and SNe~II (\verb|SNII-Templates|; \citealt{kessler2010results}), which includes both SNe~IIP/IIL; and SNe IIn (\verb|SNIIn-MOSFIT|; \citealt{villar2017theoretical}). We use the same volumetric rates for SNE~II, SNe~IIn, and SNe~Ib/c as in the PLAsTiCC challenge \citep{2015Strolger_CCSNrate}, re-scaled to match the fractional rates presented in \cite{2017Shivvers_LOSS}. The volumetric rate for SNe~Ia is taken from \cite{2018Hounsell_SNIaRate}, and that for SLSNe-I traces the star-formation history parameterized in \cite{2014Madau_SFH}. Our simulations mimic the ZTF survey strategy, filter transmissions, and reported sky noise. This results in a similar selection function favoring low-redshift ($z<0.1$) SNe as our observed sample, although we do not explicitly define a brightness threshold for photometry as is done with the BTS sample \citep{2020Fremling_ZTFBTS} and our sole quality cut is removing events with fewer than 4 total photometric observations. As a result, our simulated events are intrinsically fainter and lower-quality than our observed events.

\begin{figure}[t]
  \centering \includegraphics[width=\textwidth]{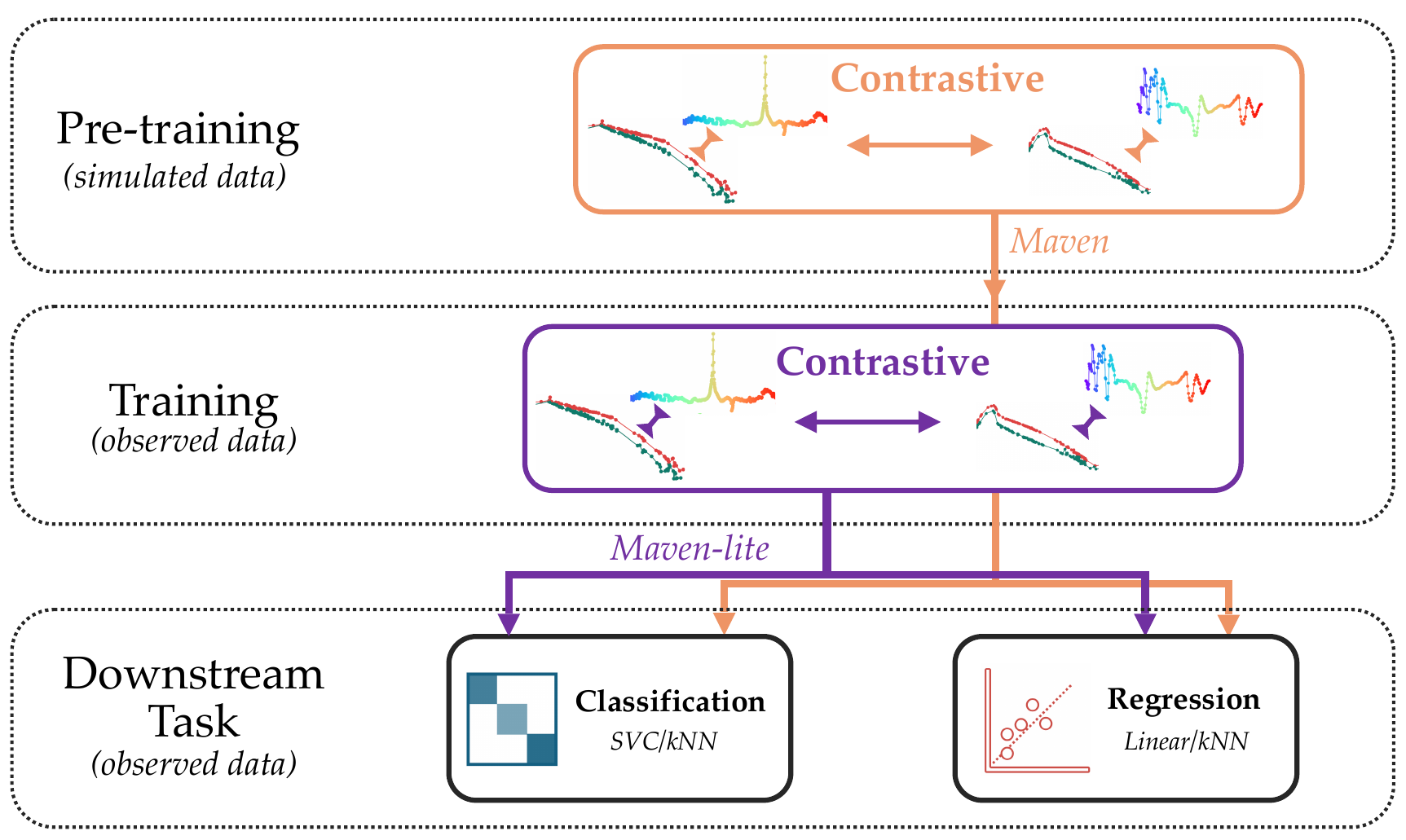} 
  \caption{{\bf Overview of our training workflows}. We first pre-train on a large simulated data set using contrastive methods (using light curves and spectra). We follow up by training on the observational ZTF dataset and then use a simple model to translate these embedding to downstream tasks. Different colors indicate different first training steps and their subsequent arrows indicate subsequent training steps.}
  \label{fig:overview_runs}
\end{figure}

In addition to the previously-developed simulations, we define a spectrograph object in SNANA with wavelength bins corresponding to the wavelength coverage of the ZTF SED machine \citep{2018Blagorodnova_SEDM}, with which the vast majority of our observed SNe were classified. To mimic the stochasticity inherent to SN classification in practice, we allow synthetic spectra to be obtained randomly from explosion to peak light, and with sufficient exposure time to achieve S/N of ~5 within an arbitrary wavelength window. Galactic extinction is applied to both modalities at the simulated SN location following the extinction law from \cite{1989Cardelli_Extinction}. 

We then pre-process all spectra in the same manner as in \citet{2019AMuthukrishna_DASH}: we apply low-pass median filtering to remove high-frequency noise, re-bin the data to log-wavelength space, and estimate the flux continuum using a polynomial fit and divide it out. While this continuum-division step removes color information, it has been shown that it has a negligible impact on redshift estimation \citep{2007Blondin_RedshiftSNe}. The spectra are kept in the observer frame (not redshift-corrected).

\subsection{The Zwicky Transient Facility Bright Transient Survey}\label{subsec:ZTFBTS}
Since 2019, the Zwicky Transient Facility (ZTF; \citealt{bellm2018zwicky}) has conducted a wide-field public survey consisting of photometry obtained with the Palomar 48-inch Schmidt telescope at a cadence of roughly 2 nights. The telescope observes in three photometric filters: ZTF-$g$, ZTF-$r$, and ZTF-$i$. Photometry is promptly reduced and streamed to alert brokers including ANTARES \citep[the Arizona-NOIRLab Temporal Analysis and Response to Events System;][]{2021Matheson_ANTARES}. For non-Galactic transients observed at or expected to peak brighter than an apparent magnitude of $\sim$18.5, a classification spectrum is automatically obtained using the Spectral Energy Distribution Machine (SEDM; \citealt{ben2012sed, blagorodnova2018sed, rigault2019fully}), a low-resolution spectrograph mounted on the Palomar 60-inch telescope \citep{cenko2006automated}. SEDM spectra are then uploaded to the Transient Name Server and the Weizmann Interactive Supernova Data Repository \citep[WISeREP;][]{2012Yaron_Wiserep}. 5,377 SNe have been spectroscopically confirmed at the time of writing as part of this Bright Transient Survey. 

We obtain metadata for 4,702 spectroscopically-classified SNe on June 18th, 2024 from the ZTF Bright Transient Survey \citep{2020Fremling_ZTFBTS} after applying all quality and purity cuts available on the ZTF BTS webpage\footnote{\url{https://sites.astro.caltech.edu/ztf/bts/bts.php}} (described in detail in \citealt{2020Perley_ZTFII}). The subsequent SNe have photometric coverage before and after peak light, good visibility throughout the photospheric phase, an uncontaminated reference image, and occurred in low extinction fields. We consolidate our resulting sample to only include events spectroscopically classified as ``normal" SN~Ia, SN~Ib/c, SN~II, SLSN-I, and SN~IIn.  

Next, we use the \texttt{Python} client of the \verb|antares| alert broker \citep{2021Matheson_ANTARES} to consolidate difference photometry for all SNe in ZTF-$g$ and ZTF-$r$ \citep[ZTF-$i$ observations are mainly private, comprising $\sim$10\% of all observations;][]{2023Aleo_YSEDR1}, and download their associated SEDM spectra from the Transient Name Server\footnote{\href{https://www.wis-tns.org/}{https://www.wis-tns.org/}} and WISEReP\footnote{\href{https://www.wiserep.org/}{https://www.wiserep.org/}} \footnote{Despite spectroscopic classifications being available on the ZTF website for all listed SNe, SEDM spectra could not be found for a few objects. When an SEDM spectrum is not available, we instead use the first reported spectrum. A positional encoding is used for the wavelengths of each spectrum, so in principle our spectrum encoder has the capacity to generalize to multiple spectrographs.}. We pre-process the observed spectra following the same procedure as our synthetic ones.

Next, we augment our observational data with noise. In each training iteration, we apply Gaussian noise to the photometric and spectroscopic observations with mean zero and standard deviation equal to the magnitude of the reported observational errors. This acts to both increase our training set and to make our model more robust to typical observational noise.

\section{Methodology}
\label{sec:methods}
\subsection{Contrastive Representation Learning}
\label{subsec:clip}

Contrastive learning is a type of self-supervised learning based on the existence of associations between data samples. It encourages corresponding data pairs to develop similar representations while separating unassociated pairs in representation space. For multimodal datasets, contrastive learning has been a common approach for aligning data pairs across modalities. Here, our goal is to build a shared representation space using photometric and spectroscopic data from the same event, and to explore the predictive properties of these representations for downstream tasks. 

For both pre-training and fine-tuning, we use the standard softmax-based bidirectional variant of the InfoNCE~\citep{oord2018representation} contrastive loss function introduced for training . Given a minibatch $\mathcal{B}$ of $|\mathcal{B}|$ associated pairs $\{(X_i, Y_i)\}_{i=1}^{|\mathcal{B}|}$ (light curves and spectra in this work), the goal is to align the learned representations of corresponding (positive) pairs $(X_i, Y_i)$ (here, the spectrum and light curve of a single SN) while repelling the representations of unaligned (negative) pairs $(X_i, Y_{j\neq i})$:

\begin{equation}
  \label{eq:softmax_loss}
  \mathcal{L}(\mathcal{B})=-\frac{1}{2|\mathcal{B}|} \sum_{i=1}^{|\mathcal{B}|}\left(\log \frac{e^{x_i \cdot y_i / \tau}}{\sum_{j=1}^{|\mathcal{B}|} e^{x_i \cdot y_j / \tau}}+\log \frac{e^{x_i \cdot y_i / \tau}}{\sum_{j=1}^{|\mathcal{B}|} e^{x_j \cdot y_i / \tau}}\right)
\end{equation}
where ${x}_i={f\left(X_i\right)}/{\left\|f\left(X_i\right)\right\|}$ and ${y}_i={g\left(Y_i\right)}/{\left\|g\left(Y_i\right)\right\|}$ are the normalized representations of the $i$-th data pairs associated with each other, and $\tau$ is a learnable hyperparameter. Encoders $f: I \rightarrow \mathbb R^{d_\text{emb}}$ and $g: T \rightarrow \mathbb R^{d_\text{emb}}$ map the two modalities to an embedding space of dimension $d_\text{emb}$. Transformer-based encoders are chosen to capture and aggregate the temporal correlations of our light curve data and the wavelength correlations of our spectroscopic data. We describe these encoders in more detail in the next section.
This loss treats the two representations symmetrically, thus ensuring that the two modalities are equally weighted. 

\begin{figure}[t]
  \centering \includegraphics[width=\textwidth]{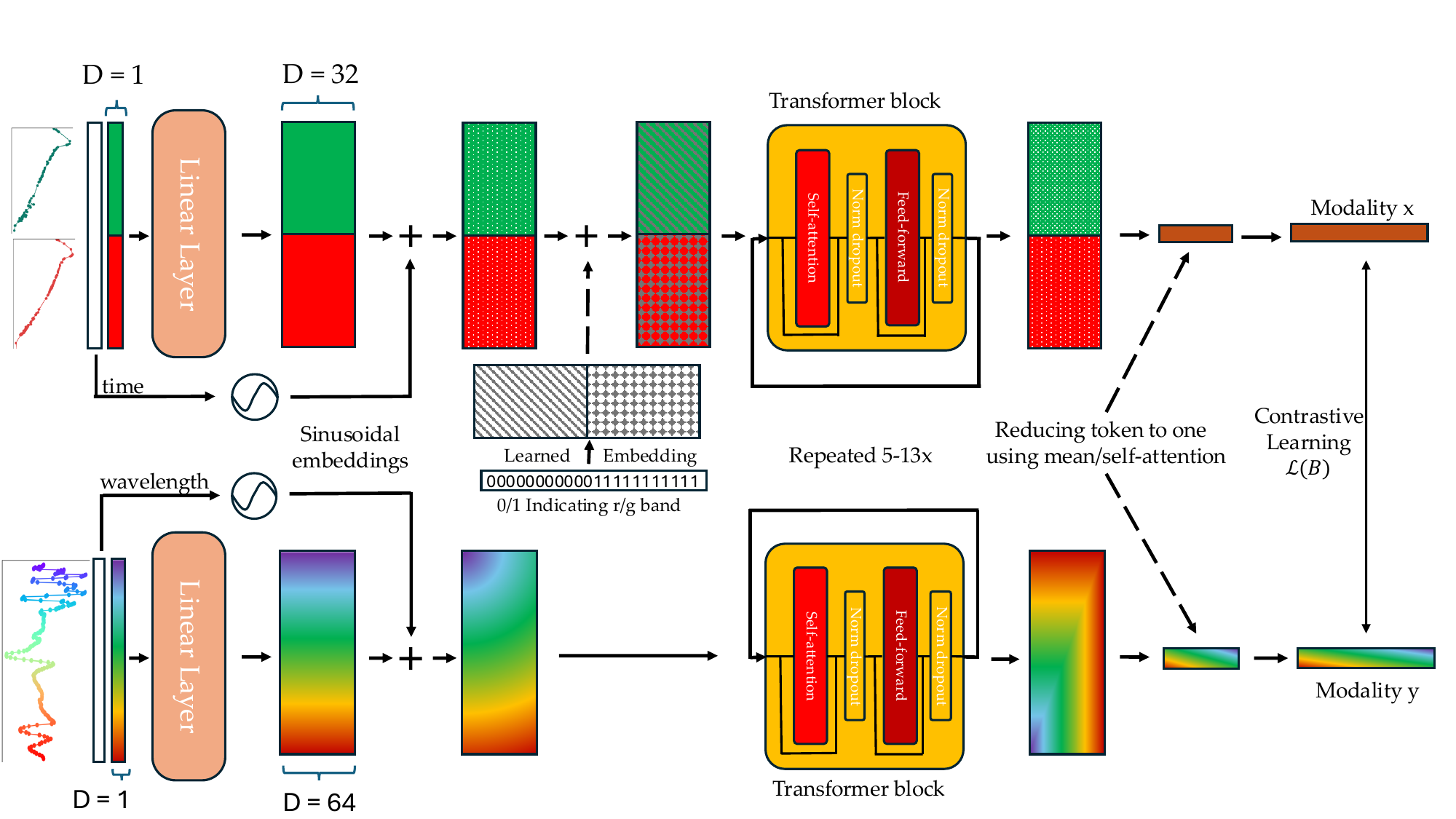} 
  \caption{{\bf Overview over model architecture}. We start by using a linear layer to increase the dimensionality of the features from one to the chosen embedding dimension. Next, we encode both time and wavelength using sinusoidal embeddings. For light curves specifically, we additionally include embeddings to represent the band information. These combined tokens are then processed through a series of transformer blocks. To reduce the output to a single token, we use a simple mean or self-attention mechanism. Finally, we project this token to match the desired length of the embedding space.}
  \label{fig:overview_transformer}
\end{figure}

\subsection{Modality Encoders}
\label{sec:encoders}

The encoders $f: I \rightarrow \mathbb R^{d_\text{emb}}$ and $g: T \rightarrow \mathbb R^{d_\text{emb}}$ are designed to efficiently extract information from high-dimensional data for the two considered modalities. Both light curve and spectrum encoders are based on the transformer architecture \citep{2017arXiv170603762V}. In this Section, we describe the architecture and explore how common representation / pre-training approaches impact downstream task performance (see Fig.~\ref{fig:overview_transformer} for an overview). 

The transformer-based light curve encoder processes magnitude measurements and their corresponding observation times. Given an input sequence of magnitude-time pairs $X = ((m_1, t_1), ..., (m_n, t_n))$, where $t_i$ is defined as the number of days from the first observation, the normalized magnitudes are initially linearly projected to the $d_\mathrm{model}$-dimensional embedding space of the transformer. Each transformer layer applies multi-head self-attention (with $n_\mathrm{heads}$ heads acting separately). Here we define
$\text{Attention}(Q, K, V) = \text{softmax}\left({QK^T}/{\sqrt{d_{k}}}\right)V$ where $Q$, $K$, and $V$ are linear projections of the input representing queries, keys, and values, and $d_{k} = d_\mathrm{model} / n_\mathrm{heads}$. A 2-layer feedforward network,
$\text{FFN}(x) = \max(0, xW_1 + b_1)W_2 + b_2$, is applied to each sequence element (a magnitude-time pair) separately. Layer normalization and residual connections are applied after attention as well as the feedforward layer.

To account for the temporal nature of light curves, we use sinusoidal time encodings to project the times $t_i$ to a higher-dimensional space,
\begin{equation}
\text{TE}(t_i, j) = \begin{cases}
\sin(t_i/n_t^{2j/d_\mathrm{model}}) & \text{if } i \text{ is even} \\
\cos(t_i/n_t^{2j/d_\mathrm{model}}) & \text{if } i \text{ is odd}
\end{cases},
\end{equation}
where $j$ is the time embedding index, $t_i$ are the input times, and $n_t$ is a hyperparameter governing the periodicity of the time encodings. This encoding allows the model to capture both absolute and relative timing of observations across a wide range of timescales.

To incorporate light curve measurements from multiple photometric filters, we concatenate all measurements for each SN and add an additional band encoding. Different bands are one-hot encoded with integers and then added to light curve magnitude embeddings before being passed into the transformer encoder. 

In contrast, the spectrum encoder processes flux measurements across multiple wavelengths. It utilizes a similar transformer-based architecture to that of the light curve encoder, but interprets the input sequence as $((f_1, \lambda_1), ..., (f_n, \lambda_n))$, where $f_i$ represents the flux at observer-frame wavelength $\lambda_i$.
The positional encoding for wavelengths follows the same sinusoidal pattern as the light curve encoder, but with $\lambda$ in place of $t$. This approach allows the model to capture both local and global spectral correlations.

For both the light curve and spectrum encoders, in addition to deterministic aggregate e.g., mean or max pooling, we consider attention-based learnable aggregation to convert the per-sequence representation to a 1-D representation vector. This enables the model to learn a data-dependent aggregation scheme, potentially better capturing correlations in the data. We initialize a learnable query vector $Q_\mathrm{learned} \in \mathbb{R}^{d_\mathrm{emb}}$, where $d_\mathrm{emb}$ is the embedding dimension. A projection of the encoded sequence after the final transformer layer, $X_\mathrm{final} \in \mathbb{R}^{n_\mathrm{seq} \times d_\mathrm{seq}}$ gives the keys and values for the attention mechanism. We use a multi-head attention architecture with two heads to then get $x_{\text{agg}} = \text{Attention}(Q_\mathrm{learned}, K_\mathrm{final}, V_\mathrm{final}) \in \mathbb R^{d_\mathrm{emb}}$ as desired. This attention-based pooling allows the model to focus on the most relevant parts of the sequence when creating the final embedding. We treat the aggregation method as a hyperparameter: in the hyperparameter tuning process discussed in section \ref{subsec:hyperpapram}, we consider both mean and attention-based aggregation.

\subsection{Transfer Learning and Fine-Tuning}
\label{subsec:finetuning}
After pre-training some of our models on the simulations discussed in section \ref{subsec:sims}, we fine-tune on the small set of ZTF BTS measurements discussed in section \ref{subsec:ZTFBTS}. 

We explore two different transfer learning approaches. First, we begin with the pre-trained model and continue training \textit{all} of its weights using the observed data. In the second approach, we again begin with the pre-trained model, but we instead allow only the weights in the first transformer block to be learnable and freeze all other weights during fine-tuning. We find that the first approach leads to better performance on downstream tasks compared to the second approach. Therefore, we only show results from the first approach hereafter. We note that both methods converged within a much smaller number of epochs than end-to-end models, despite the two-stage approach.

We define our hyperparameter-optimized pre-trained model as `Maven', and our CLIP model without pre-training as `Maven-lite' (see Fig.~\ref{fig:overview_runs}). 

\subsection{Stratified k-fold Cross-Validation}

To quantify uncertainties for both end-to-end and fine-tuned models, we perform a 5-fold cross-validation, in which we split the ZTF dataset into five unique train-test splits. All five folds share the same distribution of SN classes. The results in subsequent sections are the mean and standard deviation from these runs. To avoid added computational overhead, we do not perform it on the much larger simulation-based pre-training dataset.

\subsection{Hyperparameter Optimization}
\label{subsec:hyperpapram}
To determine hyperparameter values for model architecture and training, we perform a hyperparameter search for our end-to-end baseline and CLIP models using Weights \& Biases~\citep{wandb}. Table~\ref{tab:hyperparameter} provides a summary of the hyperparameters tuned in this process. A list of parameter values in our search are provided in configuration files in our public code repository.\footnote{\href{https://github.com/ThomasHelfer/multimodal-supernovae}{https://github.com/ThomasHelfer/multimodal-supernovae}}

In each hyperparameter sweep, we choose the set of parameter values that result in the lowest validation loss on our holdout dataset. Due to the high computational cost associated with hyperparameter tuning, we employ a random train-test split on our dataset instead of carrying out $k$-fold cross-validation. In addition, we reuse the optimal hyperparameters of the Maven-lite model for CLIP pre-training instead of performing a separate hyperparameter search. In the transfer learning stage of the pre-trained models discussed in section \ref{subsec:finetuning}, we only tune the hyperparameters shown in the rightmost column of Table~\ref{tab:hyperparameter}, while other hyperparameters are fixed to their pre-train values.

\begin{deluxetable*}{ccc}
\tablecaption{Table of hyperparameters that were optimized in a random hyperparameter search. Parameters in the leftmost column were optimized for end-to-end models while parameters in the rightmost column were optimized for all models (including end-to-end training and finetuning).\label{tab:hyperparameter}}
\tablewidth{0pt}
\tablehead{
\colhead{\bf Light curve / spectrum encoder} & 
\colhead{\bf Metadata encoder\tablenotemark{a}} & 
\colhead{\bf Optimizer and training}
}
\startdata
Number of transformer blocks & Dimension of class label embedding & Learning rate \\
Number of attention heads & Number of layers in MLP & Dropout rate \\
Normalization factor in time encoding ($n_t$) & MLP hidden layer dimension & Weight decay \\
Aggregation method & & Logit temperature in CLIP loss ($\tau$) \\
Embedding vector dimension & & Batch size \\
\enddata
\tablenotetext{a}{In addition to light curve and spectra, we explored using metadata as an additional modality. Details about the metadata encoder and results are shown in Appendix~\ref{sec:metadata}.}
\end{deluxetable*}

\subsection{Downstream Tasks}
We evaluate the performance of Maven and Maven-lite on two primary downstream tasks: classification and regression. 

Classification of SNe from photometry \textit{alone} has been an area of active study due to the long integration times required to build up sufficient signal-to-noise with spectroscopy and the subsequent rise of wide-field photometric surveys. SN classes are highly imbalanced in observed samples, due to a combination of different intrinsic volumetric rates and a steep selection function toward brighter classes (SNe~Ia). We separately consider both five-way (SN~Ia, SN~II, SN~Ib/c, SLSN-I, SN~IIn) and three-way classification (SN~Ia, SN~II, SN~Ib/c), considering in the latter case only the three most commonly-observed classes.
    
In addition to classification, we attempt to predict the redshift of each SN (which we call our ``regression" task). Redshift estimation using spectroscopic and photometric SNe~Ia is a fundamental tool for cosmological analyses. Although non-Ia classes are significantly more observationally diverse \citep[e.g.,][]{2019Modjaz_CCSNDiversity}, estimating SN redshift remains critical for estimating the intrinsic properties of an explosion (luminosity from photometry and chemical composition from spectroscopy).  

To transform our contrastive-trained light curve embeddings into classification predictions, we explore both support vector classification (SVC) and $k$-Nearest Neighbors classification ($k$NN). SVC works by finding an optimal hyperplane to separate classes. Here, we use a linear kernel with \texttt{scikit-learn} default parameters. $k$NN classification, in contrast, classifies SNe based on the similarity of their feature embedding to other latent-space neighbors.

For redshift regression, we explore both linear regression and $k$NN regression. The former uses linear transformation of the embeddings to estimate redshift, while the latter estimates redshift based on the average (or median) redshift of closest training examples in the latent space.

Lastly, we train supervised models directly on the observational ZTF dataset as our baseline models. For the classification baseline model, we optimize for the multi-class cross-entropy loss and take the class with highest pseudo-probability score as the prediction for each event in the validation set. The regression baseline model outputs a single value and is optimized using the mean squared error (MSE) loss.

\section{Results}\label{sec:results}
In this Section, we present results from Maven, Maven-lite, and our baseline models on the downstream tasks. 
\subsection{t-SNE Visualization of Latent Spaces}\label{subsec:}
\begin{figure}
    \centering
    \includegraphics[width=\linewidth]{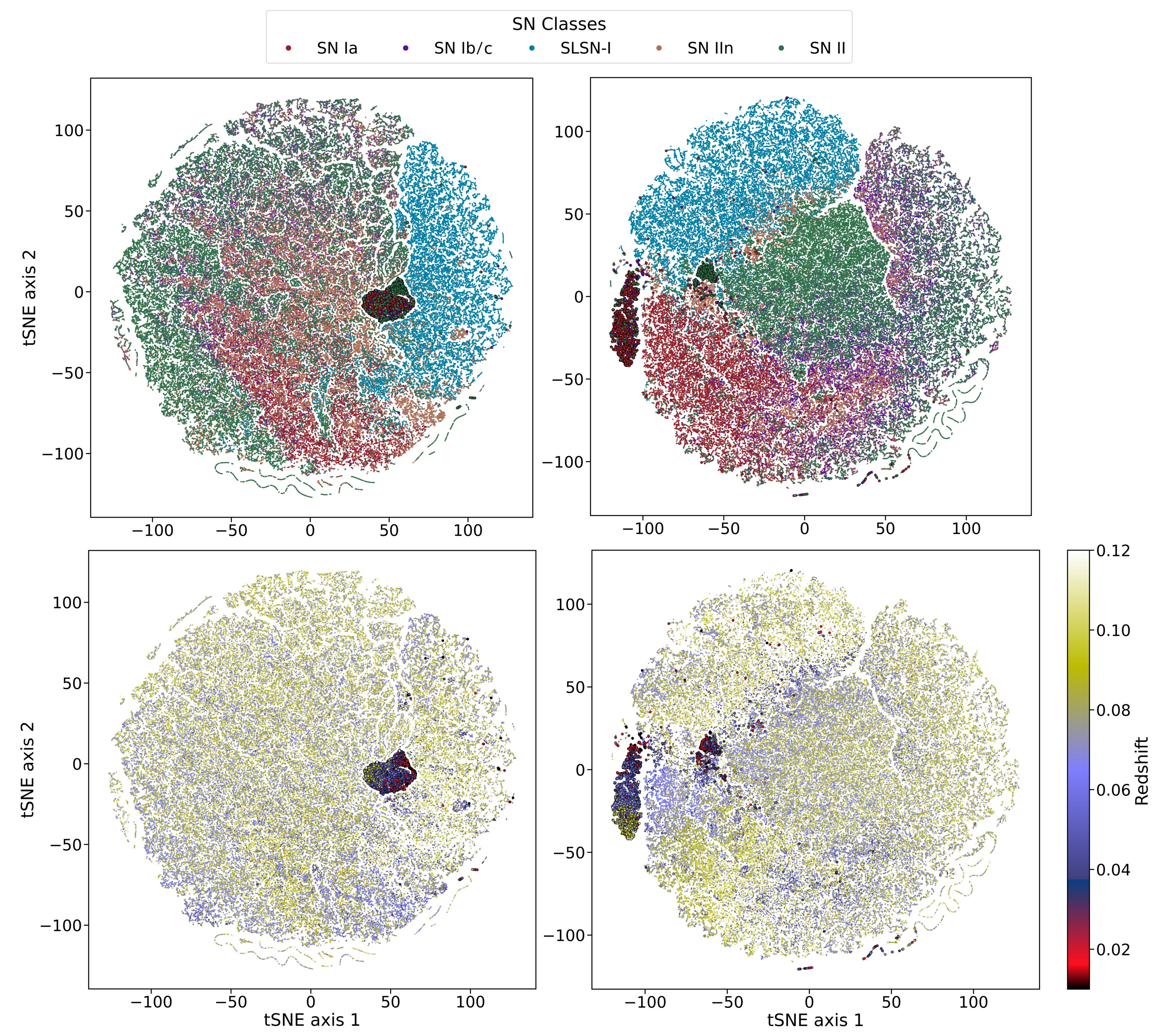}
    \caption{\textbf{Visualization of synthetic and observed light curves embedded by Maven-lite (left column) and Maven (right column).} Points in top row are colored by SN class, and points in bottom row are shaded by spectroscopic redshift. Observed data are outlined in black.}
    \label{fig:tsne_plot}
\end{figure}

To explore the impact of CLIP-style pre-training on the latent space of our Maven models, we first visualize a sample of embedded light curves. We compute Maven and Maven-lite embeddings of our five dominant classes for both the synthetic and observed samples: SNe~Ia, SNe~II, SNe~Ib/c, SLSNe~I, and SNe~IIn. We further reduce the dimensionality of our latent space using principal component analysis from the encoder output of 128 features to 50 features for computational efficiency, confirming that the subsequent 50 features retain $>$99.999\% of the variance in the original embeddings. Finally, we produce two-dimensional representations of these embeddings using the t-distributed stochastic neighbor embedding tool (t-SNE; \citealt{van2008_tsne}). Our results are presented in Fig.~\ref{fig:tsne_plot} for Maven-lite (left column) and Maven (right column), where the embeddings are colored by class in the top row and shaded by redshift in the bottom row. 

Significant differences are visible between the two latent spaces. Considering the Maven-lite embeddings, only the synthetic SLSN-I light curves (blue) are well-separated from the other classes; the core-collapse (SN~Ib/c, II, IIn) and thermonuclear (Ia) events show significant overlap. Observed Ia and II light curves (outlined in black) show similar embeddings independent of class, and little consistency with the synthetic embeddings: the majority of observed SN~Ia and SN~II lie at the boundary between synthetic SLSN-I and SN~II/SN~IIn embeddings.

In our Maven embeddings, we observe both clear separation of classes and consistent redshift gradients across our embedded light curves. The simulated SNe~Ia appear best-organized by redshift, consistent with their photometric homogeneity. The redshift gradient across observed SNe~Ia is also well-aligned with that of the synthetic sample, whereas a similar distribution is not observed in the Maven-lite embeddings. Synthetic SNe~Ib/c appear strongly mixed with both SNe~Ia and SNe~II, indicative of the photometric degeneracies between these classes.

Interestingly, although observed SN~Ia and SN~II embeddings lie closest to the synthetic events of the same class, the overlap between synthetic and observed data remains low. We attribute this to a distributional shift between synthetic and observed data. Observed events are prioritized for spectroscopic confirmation if they are brighter than (or expected to brighten above) $m<18.5$th magnitude, and additional quality and purity cuts are imposed (see Section~\ref{sec:methods} for details). While a detailed comparison between synthetic and observed events is beyond the scope of this work, this separation may also reflect the simplistic nature of our simulations relative to reality, and emphasizes the need for significantly larger observed SN samples for effective pre-training.

\newpage

\subsection{Classification Performance}\label{subsec:classificationresults}
\begin{figure}[t]
  \centering \includegraphics[width=0.83\textwidth]{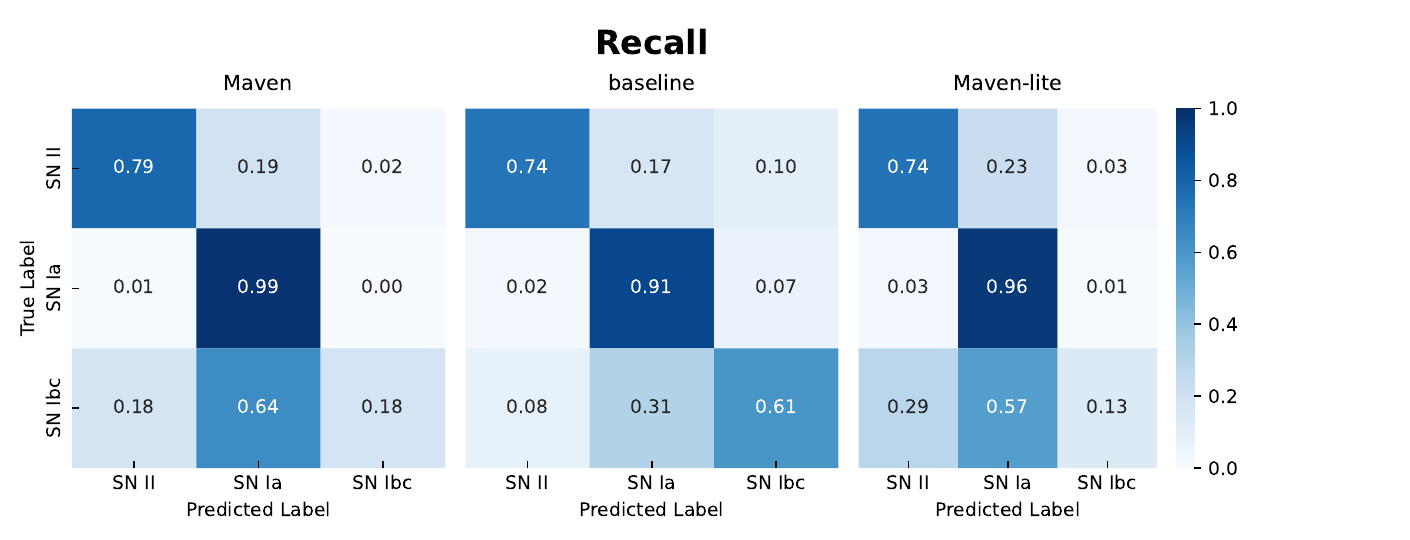} %
   \includegraphics[width=0.83\textwidth]{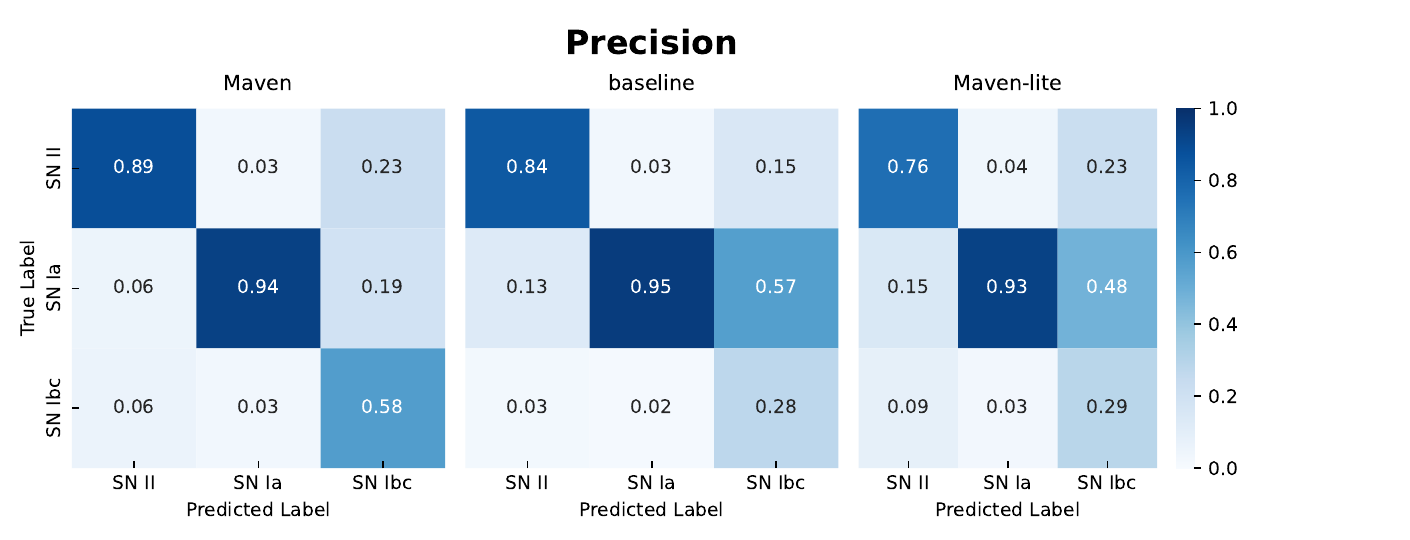}
  \caption{{\bf Normalized precision and recall confusion matrices} for supernova classification across different models and modalities. The models compared are: (a) CLIP with simulation pre-training (Maven), (b) baseline using a supervised model approach, and (c) CLIP without simulation pre-training (Maven-lite). The classes included are  SN II, SN Ia, and SN Ib/c. 
  }
  \label{fig:cm}
\end{figure}

Our results are visualized using a set of confusion matrices for our three-way classification task in Fig.~\ref{fig:cm}. We show the confusion matrices for precision (normalized by predicted class) and recall (normalized by true class) for our models. We note higher recall by Maven on the two dominant classes in our sample: 0.79 for SNe~II and 0.99 for SNe~Ia, compared to 0.74 for SNe~II and 0.91 for SNe~Ia with the baseline model. We observe poorer recall with the minority SN~Ib/c class, which comprises $\sim$5\% of the observed sample: 0.18 with simulated pre-training compared to 0.61 for the baseline. We predict that the baseline model is better able to outline the decision boundaries for this class. 

We observe the opposite results on the minority class when considering class precision. Our two-stage pre-training model achieves comparable precision to the baseline for SNe~II and SNe~Ia but substantially higher precision for SNe~Ib/c, 0.58 compared to 0.28. We note that, with substantially higher discovery rates of rare classes anticipated with the Vera C. Rubin Observatory, classification precision is essential for obtaining spectroscopic follow-up observations of relevant events. We have explored the misassociation rate as a function of event peak brightness, but identify no obvious correlations.

A common metric in classification tasks is the $F_1$ score, which increases to 1 in the limit of perfect classification. For a class $C$, $F_1$ is defined as the harmonic mean between the class's recall $r$ and precision $p$:
\begin{equation}
    F_{1,C} :  = 2\frac{p_C\times r_C}{p_C+r_C}
\end{equation}

We calculate for each model both the micro-averaged $F_1$ score, which averages performance across all events irrespective of class; and the macro-averaged $F_1$ score, which averages the $F_1$ score computed independently for each class. The macro-averaged $F_1$ score is a valuable indicator for our use case given the significant class imbalance, as the micro-$F_1$ can approach unity when all events are labeled as the dominant class. We present these results, along with the macro-averaged precision and recall (`mac-p' and `mac-r') in Table \ref{tab:main_class_3}. We further show the macro-$F_1$ score of each model as a bar plot in Fig.~\ref{fig:reg_class_f1}.

\begin{deluxetable*}{ccccccc}
\tablecaption{{\bf Overview of classification model performance.} We present three classification models: the baseline only trained on the ZTF dataset, Maven-lite without synthetic pre-training, and Maven with synthetic pretraining and observed fine-tuning. A more comprehensive overview over the runs performed in this paper can be found in Table \ref{tab:class_3_comprehensive}.}
\label{tab:main_class_3}
\tablewidth{0pt}
\tablehead{
\colhead{\bf Name} & \colhead{\bf Pre-trained} & \colhead{\bf $k$NN} & 
\colhead{\bf mac-$F_1$} & \colhead{\bf mic-$F_1$} & \colhead{\bf mac-p} & \colhead{\bf mac-r}
}
\startdata
baseline & no  & --  &  \bf 0.7011 ± 0.0303 & 0.8728 ± 0.0205 & 0.6934 ± 0.0360 & \bf 0.7527 ± 0.0247 \\
Maven & CLIP & 8     &  0.6874 ± 0.0342 & \bf 0.9247 ± 0.0070 & \bf 0.8041 ± 0.0833 & 0.6516 ± 0.0216  \\
Maven-lite & no & 3    & 0.6265 ± 0.0231 & 0.8943 ± 0.0110 & 0.6670 ± 0.0532 & 0.6119 ± 0.0121 \\
\enddata
\end{deluxetable*}

We observe macro-$F_1$ scores within 1-$\sigma$ of the baseline model for the majority of our pre-trained $k$NN classifiers, from a score of $0.6874 \pm 0.0342$ for Maven compared to a baseline of $0.7011 \pm 0.0303 $. The scores for these models are systematically higher than both Maven-lite and the majority of CLIP $k$NN classifiers without pre-training: the average $F_1$ score is 0.68 for all pre-trained $k$NN classifiers compared with an average of 0.63 for the $k$NN classifiers trained with only observed data. 

We have also calculated the performance of our models for the five-way classification task, which additionally considers the rarer classes SN~IIn and SLSN~I. Here, we observe a marginally higher average $F_1$ score for the synthetic pre-trained CLIP models than the baseline, though the results are consistent to within one standard deviation ($0.50\pm0.03$ for the best model compared to $0.49\pm0.04$). As with the three-way classification results, the macro-averaged precision of the pre-trained models is on average higher than the end-to-end baseline, with the best model achieving a score of $0.58\pm0.03$ compared to the baseline of $0.50\pm0.09$.

\subsection{Comparison to Transformer-Based SN Classifiers}

Next, we compare our multimodal model to photometric classifiers in the literature with transformer-based architectures. \cite{2024CabreraVives_ATAT} apply a custom transformer model (ATAT) to synthetic photometry and metadata from the Extended LSST Astronomical Time-Series Classification Challenge (ELAsTiCC\footnote{\url{https://portal.nersc.gov/cfs/lsst/DESC_TD_PUBLIC/ELASTICC/}}). The ATAT model consists of separate transformers, one which encodes light curves with a temporal encoding based on Fourier series and a quantile tokenizer for extracted photometric features (including the number and phases of non-detections and the flux characteristics of detections). While the model was trained and validated on synthetic photometry for 20 transient and variable astronomical classes, we can generally compare the performance by averaging the reported precision, recall, and $F_1$ scores of SNe~Ia, II, and Ib/c from their Table. Their final model achieves an average macro-$F_1$ score of 0.67 across the three classes, compared with 0.70 for our end-to-end baseline and 0.69 for our best-performing light curve and spectra-aligned models. They report an average recall of 0.71 for these classes, compared to 0.75 for our end-to-end baseline and 0.65 for Maven; and an average precision of 0.63, compared to 0.69 for our baseline and 0.80 for Maven. We caution that these datasets are distinct, limiting further comparison.

The results of \cite{2023Pimentel_TransformerClassification} are more directly comparable to this work. \cite{2023Pimentel_TransformerClassification} present a transformer model for ZTF photometry in which the time of each observation is encoded as the phase from first detection using a Fourier decomposition-based temporal modulation, with noise added to the values in training to prevent overfitting. In a two-stage pre-training scheme with both synthetic and observed events, the optimization problem is defined with reconstruction and cross-entropy regularization terms to preserve class-specific information in the encoded light curves. The resulting `S-TimeModAttn' model is trained and validated on $g$ and $r$-band light curves from the Zwicky Transient Facility, with presumably substantial overlap with the observational dataset considered in this work. \cite{2023Pimentel_TransformerClassification} report a macro-$F_1$ of $0.614\pm0.036$ in the task of four-way classification (Ia, II, Ib/c, and SLSN), compared to our $0.6874 \pm 0.0342$ for three-way classification; and a macro-precision of $0.598\pm0.030$ compared to our $0.804\pm0.083$. A macro-recall (also referred to as completeness) score of 0.72 for three-way classification can be inferred from their confusion matrices in Fig.~\ref{fig:cm}, compared to our lower $0.6516\pm0.0216$. Class-specific $F_1$ scores and precisions (also referred to as purity) are not reported.

\subsection{Regression Performance}\label{subsec:regressionresults}

\begin{figure}[t]
  \centering \includegraphics[width=1.0\textwidth]{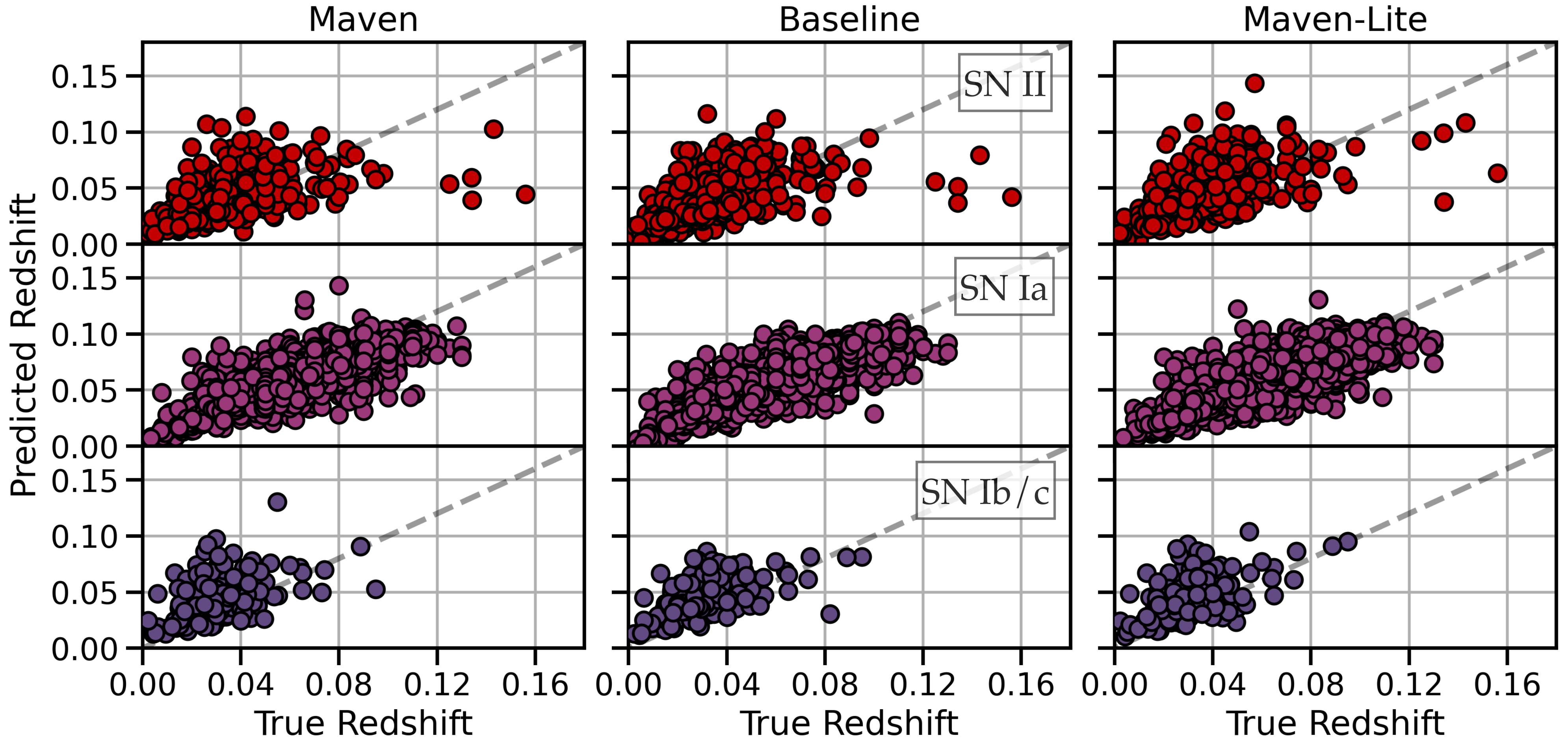}  
    \caption{{\bf Predicted versus true redshift} for three different models: our pre-trained Maven (left), a baseline model using supervised methods, and our Maven-lite model trained end-to-end on the observed ZTF data. The data is segregated into different SN classes. The transparent points in the background represent all other classes. The SNe classes are displayed in separate columns, and the models are shown in separate rows for clear comparison.}
  \label{fig:predvstrueredshift}
\end{figure}

\begin{figure}[h!]
  \centering \includegraphics[width=0.47\textwidth]{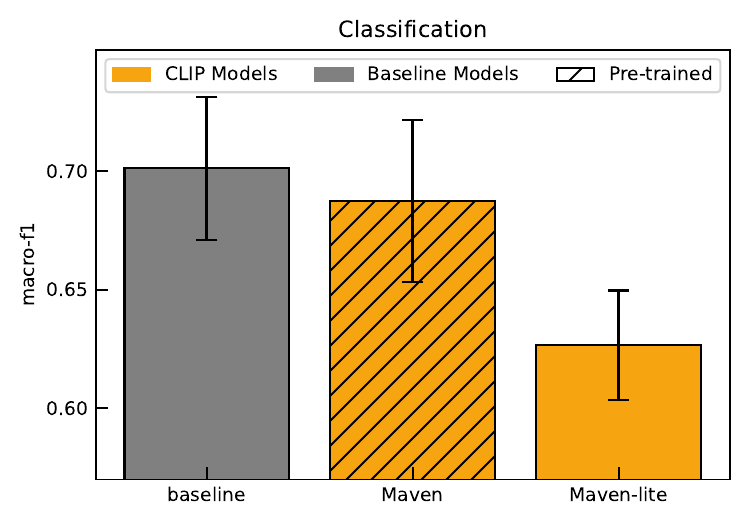} %
   \includegraphics[width=0.47\textwidth]{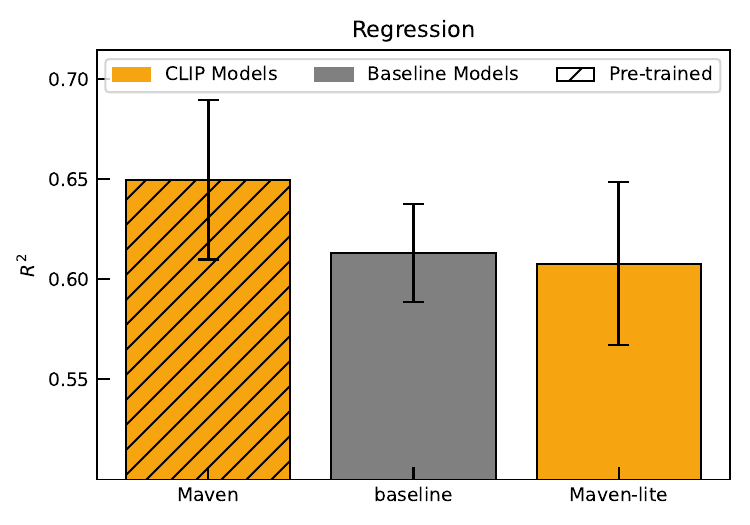}
  \caption{{\bf Final performance metrics} for Maven, Maven-lite, and baseline models for on downstream classification and regression tasks.}
  \label{fig:reg_class_f1}
\end{figure}

We next consider the task of redshift estimation. We quantify the performance of our models with the coefficient of determination $R^2$, the L1 and L2 error, and the outlier fraction `OLF', defined as $|z_{\rm{pred}} - z_{\rm{true}}|/(1+z_{\rm{true}}) > 0.15$. We report these values for contrastive pre-trained models in Table~\ref{tab:main_regression}. We also present a bar plot of the R$^2$ values in Fig.~\ref{fig:reg_class_f1}, and the predicted versus true redshifts for each SN class in Fig.~\ref{fig:predvstrueredshift}. As expected, we observe the highest correlation between observed and predicted redshifts for SNe~Ia, the most observationally homogeneous SN class considered. We calculate an $R^2$ value of $R^2=0.6496 \pm 0.0398 $ for Maven compared to the end-to-end baseline performance of $R^2=0.6129 \pm 0.0245$. The L1 and L2 errors are also lower on average for Maven than for our regression baseline, while the outlier fraction is comparable. We conclude that, on average, Maven outperforms the baseline regression model. Maven-lite, our model without pre-training, achieves an $R^2$ value of $0.6078\pm0.0408$, lower than both Maven and the baseline model.

Though a comparable photometric redshift model for low-redshift ZTF SNe does not exist in literature, an outlier fraction of 0.004 is reported for 289 photometric SNe~Ia in the Supernova Legacy Survey (SNLS), nearly an order of magnitude higher than our best model but with a substantially higher maximum redshift $z<1.0$ \citep{2010Palanque_photozs}. Another analytic photometric redshift estimator proposed by \cite{2015Wang_LSST} for SNe~Ia discovered by LSST finds an outlier fraction of 0.0023 over $z<1.0$, compared to our 0.0002.

\begin{deluxetable*}{ccccccc}[!t]
\tablecaption{{\bf Overview of regression model performance.} We present three regression models: the baseline only trained on the ZTF dataset, a CLIP model trained only on the ZTF dataset (Maven-lite) and a CLIP model pre-trained on simulated data (Maven) and then subsequently trained on ZTF. A more comprehensive overview over the runs performed in this paper can be found in Table \ref{tab:regress_comprehensive}.
\label{tab:main_regression}}
\tablehead{
\colhead{\bf Name} & \colhead{\bf Pre-trained} & \colhead{\bf $k$NN} & \colhead{\bf R2} & \colhead{\bf L1} & \colhead{\bf L2} & \colhead{\bf OLF}
}
\startdata
Maven & CLIP & 9 & \bf  0.6496 ± 0.0398 & \bf 0.0095 ± 0.0004 & \bf 0.0152 ± 0.0014 & 0.0002 ± 0.0005 \\
baseline & no &  & 0.6129 ± 0.0245 & 0.0104 ± 0.0004 & 0.0160 ± 0.0010 & 0.0002 ± 0.0005 \\
Maven-lite & no & 9 & 0.6078 ± 0.0408 & 0.0103 ± 0.0006 & 0.0161 ± 0.0014 & 0.0002 ± 0.0005 \\
\enddata
\end{deluxetable*}

\section{Discussion}\label{sec:discussion}

We have explored the value of constrastive pre-training in constructing a foundational model for SN science. By first training with synthetic events and fine-tuning with observed events, we have constructed a model, Maven, whose performance on the downstream tasks of photometric classification and redshift is on par with models optimized end-to-end for these tasks. Maven outperforms our classification baseline model, with a micro-averaged-$F1$ score of 0.92. Similarly, Maven outperforms our baseline for redshift regression, with an L2-loss of $0.015$ and minimal outlier fraction. While we have limited our study to ZTF data, adapting Maven to incorporate additional photometric filters and classes of astronomical transients would allow us to repurpose it for diverse time-domain studies with the Vera C. Rubin Observatory using fewer computational resources than building multiple specialized models. 

CLIP-style pre-training has been proposed as a simple and effective mechanism for extracting information from multiple modalities in a single model. The following conditions need to be met for multimodal constrastive learning to be effective: that significant information content is shared across these modalities; that the mutual information is relevant for the downstream tasks; and that the shared information is the maximal information in each modality relevant for the downstream tasks. Recently, \cite{2023Liang_MultiView} formalized this picture by defining `multi-view redundancy' as a necessary condition for effective pre-training using traditional constrastive learning. In our case, we know spectra to be highly informative for both classification (the taxonomy is \textit{defined} by spectra obtained early after a SN's explosion, with the temporal evolution of the explosion rarely considered) and redshift inference, which is achieved primarily through the identification of spectral lines. Supernova photometry, although containing some spectral information, is significantly more lossy: the collection of photons through a broadband photometric filter destroys valuable information about a supernova's underlying spectral energy distribution that might otherwise be valuable for these tasks (as seen in the diagram in Fig.~\ref{fig:overview_SED}). For these reasons, we can characterize supernova light curves as an `information-poor' modality and spectra as an `information-rich' modality for our tasks. CLIP-style pre-training, in this case, is unable to bring significant performance gains beyond end-to-end optimized models. This behavior persists despite aligning these modalities directly with the relevant downstream information (metadata of an event's spectroscopic classification and redshift, as discussed in Appendix~\ref{sec:metadata}): \emph{the least-informative modality sets an upper limit on the mutual information that can be extracted.}

Seen from this perspective, it is surprising that we do not observe substantial \textit{drops} in performance relative to our baseline models. We attribute this to systematic hyperparameter tuning and additional synthetic pre-training, through which we are able to mitigate the negative effects of constrastive alignment. We therefore advise caution in the use of multimodal constrastive pre-training, which should be specialized for the input modalities and the anticipated downstream tasks. In our case, additional improvements may be possible with a pre-training scheme designed to preserve both mutual and unique information content relevant for classification and redshift estimation (as is proposed in \citealt{2023Liang_MultiView}). 

\section{Conclusion}\label{sec:conclusion}
We conclude by summarizing our key findings:
\begin{enumerate}
    \item We train Maven through self-supervised contrastive learning on SN spectra and light curves. Maven is able to achieve state-of-the-art performance on both redshift estimation and SN classification. 
    \item We find that pre-training on a large simulated dataset significantly improves Maven's performance on downstream tasks over a contrastively-trained model on solely the observed data. 
    \item Maven does not dramatically outperform supervised models optimized directly for each downstream task. We hypothesize that this is due to the light curve being an information-poor modality, which limits the amount of information our unsupervised objective is able to extract. 
\end{enumerate}

Starting in 2025, the Vera C. Rubin Observatory will initiate the 10-year Legacy Survey for Space and Time, and detect $\sim$1M SNe yr$^{-1}$ in \textit{ugrizY} filters. This consistently-calibrated photometric dataset will enable self-supervised pre-training for time-domain foundation models (including variable stars, lensing events, active galactic nuclei, and other non-SN phenomena) at an unprecedented scale. However, without spectroscopy for the vast majority of detected events, the self-supervised tasks that can be applied with this data will be limited to a single modality.

On the other hand, traditional multimodal models have considered distinct representations of a single astronomical object (photometry and spectroscopy of a supernova). Where spectroscopic \textit{and} photometric information for a transient is sparse, however, broad physical properties can be inferred from the event's host galaxy \citep{2012Hakobyan_HostGalaxies,2020Kang_IaHosts,2021Schulze_PTF,2024Chakraborty_IaHosts}. Early efforts have emphasized the value of these data for photometric classification \citep{2020Gomez_FLEET,2021Carrasco_StampClassifier, 2023Gagliano_FirstImpressions,2024Sheng_NEEDLE}. LSST data will contain photometry for tens of billions of galaxies, millions of which will be spectroscopically-confirmed through the Dark Energy Spectroscopic Instrument \citep[DESI;][]{2019Levi_DESI} or 4MOST \citep{2023Dumayne_4MOST}. Additional work should be dedicated to exploring the linking of modalities spanning distinct lengthscales, which would allow for both supernova and host-galaxy data to be consolidated in a single pre-training scheme.

\begin{acknowledgments}
    This work was initiated at the IAIFI AstroML Hackathon, held at MIT in January 2024.
    This work is supported by the National Science Foundation under Cooperative Agreement PHY-2019786 (The NSF AI Institute for Artificial Intelligence and Fundamental Interactions, \url{http://iaifi.org/}) and AST-2108676. This material is based upon work supported by the U.S. Department of Energy, Office of Science, Office of High Energy Physics of U.S. Department of Energy under grant Contract Number  DE-SC0012567. 
    This work was performed in part at the Aspen Center for Physics, which is supported by National Science Foundation grant PHY-2210452.
    Research reported in this publication was supported by a Postdoctoral Fellowship at the Institute for Advanced Computational Science, Stony Brook University. This work was carried out at the Advanced Research Computing at Hopkins (ARCH) core facility  (rockfish.jhu.edu), which is supported by the National Science Foundation (NSF) grant number OAC1920103, as well as the FASRC Cannon cluster supported by the FAS Division of Science Research Computing Group at Harvard University. 
\end{acknowledgments}

\software{
   \package{Jupyter} \citep{Kluyver2016jupyter}, \package{Matplotlib} \citep{Hunter:2007}, \package{Numpy} \citep{harris2020array}, PyTorch \citep{NEURIPS2019_9015}, \package{PyTorch lightning} \citep{Falcon_PyTorch_Lightning_2019}, 
   \package{Astropy} \citep{astropy:2013, astropy:2018, astropy:2022},
   \package{Einops}\citep{rogozhnikov2022einops}
   \package{Pandas}\citep{reback2020pandas}, \package{scikit-learn}\citep{scikit-learn} and \package{wandb} \citep{wandb}.
    }

\bibliography{multimodal}
\bibliographystyle{aasjournal-mod}

\appendix

\section{Metadata CLIP}\label{sec:metadata}
In addition to SN spectrum and light curve measurements, we also considered SN metadata as an additional modality for training a CLIP model. The metadata modality used in our training includes supernovae redshifts and class labels. We encode each class label with a learnable embedding vector. The metadata encoder consists of a multilayer perceptron~(MLP) that takes in the concatenated vector of class embedding and redshift and outputs the final embedding. The number of hidden layers and the hidden layer dimension in the MLP were chosen from a hyperparameter search, as discussed in Sec~\ref{subsec:hyperpapram}.

The models which directly align event photometry with relevant metadata (redshift and class) in pre-training do not significantly outperform the models in which photometry and spectroscopy alone are aligned. Considering only pre-trained models for the task of classification, we observe comparable three-way macro-$F_1$ scores when aligning light curves with metadata ($0.692\pm0.022$), light curves with spectra ($0.687\pm0.034$), and light curves with both spectra and metadata ($0.685\pm0.019$). Each of our CLIP objectives featured photometry as a modality, and we predict that this more information-poor modality is driving the observed performance across each of these models, as we discuss in additional detail in section \ref{sec:conclusion}.

\begin{deluxetable*}{cccccc}[!h]
\tablecaption{{\bf Classification performance for three classes by model configuration }: This table presents the classification performance of various models using light curve data from the ZTF dataset. The models are categorized based on whether they utilized simulation pre-training (`pre-trained'), the type of last layer added to embedding models (`last-layer'). The modalities taken into account when training on the real ZTF dataset are indicated in `real-pre' (lc - light curve, sp - spectrum, m - metadata) as well as whether a SVC or $k$NN. Performance metrics include macro-F1 (mac-f1), micro-F1 (mic-f1), macro-precision (mac-p), and macro-recall (mac-r). The results are presented as mean ± standard deviation, calculated over five folds. Baseline models, trained in an end-to-end supervised fashion using only the ZTF data, are included for comparison.}\label{tab:class_3_comprehensive}
\tablehead{
\colhead{\textbf{pre-trained}} & \colhead{\textbf{last-layer}} & \colhead{\textbf{real-pre}} & \colhead{\textbf{mac-f1}} & \colhead{\textbf{mac-p}} & \colhead{\textbf{mac-r}}
}
\startdata
no & \multicolumn{2}{c}{end-to-end baseline}  & 0.7011 ± 0.0303 & 0.6934 ± 0.0360 & 0.7527 ± 0.0247 \\
clip & $k$NN & lc-m &   0.6920 ± 0.0217 & 0.7286 ± 0.0377 & 0.6721 ± 0.0183 \\
clip & $k$NN & lc-sp &    0.6874 ± 0.0342 & 0.8041 ± 0.0833 & 0.6516 ± 0.0216 \\
clip & $k$NN & lc-sp-m &    0.6849 ± 0.0194 & 0.7280 ± 0.0334 & 0.6643 ± 0.0161 \\
clip & SVC & lc-m &    0.6747 ± 0.0297 & 0.8026 ± 0.0257 & 0.6435 ± 0.0257 \\
clip & SVC & lc-sp-m &  0.6522 ± 0.0237 & 0.7892 ± 0.0975 & 0.6247 ± 0.0215 \\
no & $k$NN & lc-sp-m &    0.6268 ± 0.0251 & 0.7204 ± 0.0701 & 0.6000 ± 0.0199 \\
no & $k$NN & lc-sp &  0.6265 ± 0.0231 & 0.6670 ± 0.0532 & 0.6119 ± 0.0121 \\
no & $k$NN  & lc-m &   0.6249 ± 0.0228 & 0.7309 ± 0.0661 & 0.6035 ± 0.0184 \\
clip & SVC & lc-sp &   0.6195 ± 0.0190 & 0.7753 ± 0.0994 & 0.6056 ± 0.0172 \\
no & SVC & lc-m &  0.5971 ± 0.0220 & 0.7871 ± 0.1858 & 0.5842 ± 0.0163 \\
no & SVC & lc-sp-m &  0.5938 ± 0.0156 & 0.7892 ± 0.1873 & 0.5802 ± 0.0077 \\
no & SVC & lc-sp & 0.5749 ± 0.0099 & 0.5857 ± 0.0126 & 0.5686 ± 0.0102 \\
\enddata
\end{deluxetable*}

\begin{deluxetable*}{cccccc}[!h]
\tablecaption{{\bf Regression Performance by Model Configuration}: This table presents the regression performance of various models using light curve data from the ZTF dataset. The models are categorized based on whether they utilized simulation pre-training (`pre-trained'), the type of last layer added to embedding models (`last-layer'). The modalities taken into account when training on the real ZTF dataset is indicated in `real-pre' (lc - light curve, sp - spectrum, m - metadata) as well weather we use a linear or $k$NN layer to translate our embedding to a redshift prediction (`last-layer`). Performance metrics include the coefficient of determination ($R^2$), L1 loss, and L2 loss. The results are presented as mean ± standard deviation, calculated over five folds. Baseline models, trained in an end-to-end supervised fashion using only the ZTF data, are included for comparison.}
\label{tab:regress_comprehensive}
\tablewidth{0pt}
\tablehead{\colhead{\bf pre-trained} & \colhead{\bf last-layer} & \colhead{\bf real-pre} & \colhead{\bf R2} & \colhead{\bf L1} & \colhead{\bf L2}}
\startdata
clip & $k$NN & lc-m &       0.6543 ± 0.0280 & 0.0094 ± 0.0005 & 0.0152 ± 0.0010 \\
clip & Linear & lc-sp-m &       0.6513 ± 0.0440 & 0.0096 ± 0.0005 & 0.0152 ± 0.0016 \\
clip & $k$NN & lc-sp &       0.6496 ± 0.0398 & 0.0095 ± 0.0004 & 0.0152 ± 0.0014 \\
clip & $k$NN & lc-sp-m &       0.6470 ± 0.0422 & 0.0094 ± 0.0006 & 0.0152 ± 0.0012 \\
clip & Linear & lc-sp &       0.6386 ± 0.0447 & 0.0099 ± 0.0003 & 0.0155 ± 0.0016 \\
clip & Linear & lc-m &       0.6345 ± 0.0444 & 0.0100 ± 0.0006 & 0.0156 ± 0.0014 \\
no & $k$NN &  lc-m &       0.6150 ± 0.0294 & 0.0103 ± 0.0003 & 0.0160 ± 0.0012 \\
no & \multicolumn{2}{c}{end-to-end baseline}   & 0.6129 ± 0.0245 & 0.0104 ± 0.0004 & 0.0160 ± 0.0010 \\
no & $k$NN & lc-sp-m &       0.6090 ± 0.0464 & 0.0102 ± 0.0005 & 0.0161 ± 0.0015 \\
no & $k$NN & lc-sp &       0.6078 ± 0.0408 & 0.0103 ± 0.0006 & 0.0161 ± 0.0014 \\
no & Linear & lc-sp &       0.5948 ± 0.0402 & 0.0107 ± 0.0007 & 0.0164 ± 0.0015 \\
no & Linear & lc-sp-m &       0.5938 ± 0.0450 & 0.0108 ± 0.0004 & 0.0164 ± 0.0016 \\
no & Linear & lc-m &      0.5927 ± 0.0399 & 0.0107 ± 0.0004 & 0.0165 ± 0.0015 \\
\enddata
\end{deluxetable*}

\end{document}